\documentclass[12pt]{article}
\usepackage{amsmath,amssymb,amsfonts,amsthm}

\title{Quantum entanglement and position-momentum entropic squeezing of a moving Lambda-type three-level atom interacting with a single-mode quantized field with intensity-dependent coupling}

\author{M J Faghihi$^{1,3,4}$ and M K Tavassoly$^{1,2,3,*}$ \\
 \footnotesize{$^1$ Atomic and Molecular Group, Faculty of Physics, Yazd University, Yazd, Iran} \\
 \footnotesize{$^2$ Photonics Research Group, Engineering Research Center, Yazd University, Yazd, Iran} \\
 \footnotesize{$^4$ The Laboratory of Quantum Information Processing, Yazd University, Yazd, Iran} \\
 \footnotesize{$^5$ Physics and Photonics Department, Graduate University of Advanced Technology, Mahan, Kerman, Iran} \\
 \footnotesize{$^*$ E-mail: mktavassoly@yazd.ac.ir}}

\begin{document}
\maketitle
%=======================================================================
%=======================================
 %=======================================

 \newcommand{\norm}[1]{\left\Vert#1\right\Vert}
 \newcommand{\abs}[1]{\left\vert#1\right\vert}
 \newcommand{\set}[1]{\left\{#1\right\}}
 \newcommand{\R}{\mathbb R}
 \newcommand{\I}{\mathbb{I}}
 \newcommand{\C}{\mathbb C}
 \newcommand{\eps}{\varepsilon}
 \newcommand{\To}{\longrightarrow}
 \newcommand{\BX}{\mathbf{B}(X)}
 \newcommand{\HH}{\mathfrak{H}}
 \newcommand{\A}{\mathcal{A}}
 \newcommand{\D}{\mathcal{D}}
 \newcommand{\N}{\mathcal{N}}
 \newcommand{\x}{\mathcal{x}}
 \newcommand{\p}{\mathcal{p}}
 \newcommand{\la}{\lambda}
 \newcommand{\af}{a^{ }_F}
 \newcommand{\afd}{a^\dag_F}
 \newcommand{\afy}{a^{ }_{F^{-1}}}
 \newcommand{\afdy}{a^\dag_{F^{-1}}}
 \newcommand{\fn}{\phi^{ }_n}
 \newcommand{\HD}{\hat{\mathcal{H}}}
 \newcommand{\HDD}{\mathcal{H}}
%==================================================

 \begin{abstract}
 In this paper, we study the interaction between a moving $\Lambda$-type three-level atom and a single-mode cavity field in the presence of intensity-dependent atom-field coupling. After obtaining the state vector of the entire system explicitly, we study the nonclassical features of the system such as quantum entanglement, position-momentum entropic squeezing, quadrature squeezing and sub-Poissonian statistics. According to the obtained numerical results we illustrate that the squeezed period, the duration of entropy squeezing and the maximal squeezing can be controlled by choosing the appropriate nonlinearity function together with entering the atomic motion effect by suitably selection of the field-mode structure parameter. Also, the atomic motion, as well as the nonlinearity function leads to the oscillatory behaviour of the degree of entanglement between the atom and field.
 \end{abstract}

%==============================================
 %==============================================

 %==========================================================================
 \section{Introduction}\label{sec-intro}
 %==========================================================================

 Quantum entanglement is one of the most profound features and prominent trait of quantum mechanics that has been attracted a lot of attention.
 Also, it has been considered to be a worth physical resource in the quantum information science contains quantum computation and communication \cite{b1}, quantum dense coding \cite{b2}, quantum teleportation \cite{b3}, quantum information processing \cite{b4}, sensitive measurements \cite{sm} and so on.
 It is valuable to mention that, in quantum information processing, generating and manipulating the quantum entangled states, is one of the key problems that should be dispelled before paying attention to anything else. A simple way that realizes this aim and produces the quantum entangled state is the atom-field interaction in a cavity QED, which is led to the atom-field entangled state. \\
 The fully quantum mechanical description of the atom-field interaction is achieved by appropriately generalizing the Jaynes-Cummings model (JCM) \cite{JCM} which is a simplified version of the interaction of the single-mode field and the two-level atom in the rotating wave approximation (RWA). This model is a mandatory subject in quantum optics and so has been generalized in the literature. For instance, intensity-dependent JCM has been suggested by Buck and Sukumar \cite{suk} and then has been used by others \cite{selective}.
 Quantum properties of a $V$-type and $\Lambda$-type three-level atom interacting with a single-mode field in a Kerr medium with intensity dependent coupling and in the presence of the detuning parameters have been studied in \cite{zait} and \cite{us} by Zait and us, respectively.

 Abdallah {\it et al} \cite{abd1} considered a model, in which two fields interact with a two-level atom within perfect cavity, in addition to the presence of the field-field interaction such that the fields are assumed to be in the parametric amplifier form. Then, after calculating the atomic occupation probabilities, they analyzed the degree of entanglement as well as the phase distribution function for their model.
 The generalized JCM which consists of a two-level atom interacting with two modes of the electromagnetic radiation field including the field-field interaction (frequency conversion) has been introduced by Khalil {\it et al} \cite{abd2}. The time evolution of the second-order correlation function in the interaction of a three-level atom with a single mode of the quantized electromagnetic cavity field has been discussed by Abdel-Wahab \cite{wahab}.
  The JCM in the presence of the  action of an external classical field has been studied by  Abdalla {\it et al} \cite{abd3}.
 Jin-Liang Guo {\it et al} \cite{guo} examined individually the influences of the Kerr-like medium and intensity dependent coupling on entropy exchange and entanglement in the JCM and showed that these effects are helpful for improving the quality of entropy exchange. A general formalism for a $\Lambda$-type three-level atom interacting with a correlated two-mode field has been presented in \cite{aty}, in which the authors found the degree of entanglement for their system by using the density matrix operator approach. Also, a model for the interaction of a three-level atom in the $\Lambda$-configuration with a two-mode field under a multi-photon process has been recently introduced by Obada {\it et al} \cite{obada}, in which they studied the effects of photon-number, detuning and nonlinearities of both the field and intensity-dependent atom-field couplings on the degree of entanglement. \\
 On the other hand, due to the fact that in any atom-field interaction, the atom may not be performed to be exactly static during the interaction, the effect of atomic motion on the interaction dynamics should be taken into account.
 For instance, the influence of atomic motion and field-mode structure on the atomic dynamics (atomic population inversion) has been examined by Joshi {\it et al} \cite{joshi1}. Also, a model in which a moving atom undergoes a two-photon transition in a two-mode coherent state field has been studied by Joshi \cite{joshi2}. The authors then compared their own results with those of \cite{schlicher} by Schlicher in, which an atom undergoes a one-photon transition. Liao {\it et al} \cite{liao1} investigated the entropy squeezing of a moving atom interacting with a single-mode quantized field and investigated some effects on the evolution of this property. Also, quantum entanglement of the SU(1,1)-related coherent state which interacts with a moving atom has been outlined in \cite{liao2}. \\
 As another point, it is recently found that in quantum information studies, three-level systems possess the outstanding advantages in comparison with two-level systems \cite{helle}. Bru{\ss} {\it et al} \cite{brub} have shown that quantum key distribution schemes in quantum cryptography with three-level systems are more secure against symmetric attacks than protocols based on two-dimensional quantum variables (qubits). Kaszlikowski {\it et al} \cite{Kaszlikowski} have generalized the two entangled quantum systems to $N$-dimensional Hilbert spaces or $` `$quNits$"$ and demonstrated that based on Bell's inequality \cite{bell}, quantum nonlocality in three-level systems ($N=3$) is stronger than that in two-level systems. Thus, the three-level or even multilevel systems are now of more interest in quantum information processing. For instance, one can pay attention to the conflict of local realism and quantum mechanics, by recalling the Greenberger-Horne-Zeilinger (GHZ) theorem \cite{ghz}. According to this theorem, for three or more qubits this conflict is much sharper than for two qubits. In addition, there are some suggestions which indicate that the conflict of local realism and quantum mechanics diminishes with growing N (dimension of Hilbert space) \cite{growingN}. So, it seems that three- or multi-level atoms which may be considered as three or more qubits can be better candidates to utilize in the theoretical/experimental observations. This is one of the reasons for which three-level atoms have received remarkable attention in the literature which concerns with the atom-field interaction \cite{us,3level}. In addition, among the known three-level atomic configurations, the advantages of $\Lambda$-configuration are reported in comparison with $V$-type atom, in the sense of showing the nonclassicality features.
 We have newly shown that the nonclassical properties (particularly, the degree of entanglement between the atom and field, sub-Poissonian statistics and different orders of squeezing) are more visible in $\Lambda$-type, in comparison with $V$-type three-level atoms \cite{us}.
 In addition, Civitarese {\it et al} \cite{Civitarese} have found that atomic squeezing becomes evident in both ladder and $\Lambda$ schemes of three-level atoms.
 Also, they demonstrated that, regardless of the choice of the coupling constants and the number of atoms and photons, spin squeezing does not appear so clearly in $V$-type three-level atoms. \\
 Adding the above discussion, in this paper we intend to study a $\Lambda$-type three-level atom in motion which interacts with a single-mode cavity field in the presence of intensity-dependent coupling. Briefly speaking, the main goals of the present paper is to discuss the effects of intensity-dependent atom-field coupling together with the atomic motion as well as the field-mode structure parameter on the nonclassical features of the state vector of the whole system such as quantum entanglement (time evolution of the field entropy), entropy squeezing (by using the position-momentum entropic uncertainty relation), photon statistics and normal squeezing of the quadratures of the field.\\
 This paper is organized as follows: In the next section, we obtain the state vector of the whole system using the generalized JCM. In section 3, quantum entanglement due to the atom-field interaction is evaluated and entropy squeezing is studied in section 4. Then, we pay attention to the quantum statistics of the system by considering the Mandel parameter in section 5 and section 6 deals with the normal squeezing of the field quadratures. Finally, section 7 contains a summary and concluding remarks.

 %==================================================================================
 \section{Introducing the model and its solution}
 %==================================================================================

 In quantum mechanics, possible information in studying any physical system arises from the wavefunction of the system. Now, we plan to allocate this section to obtain the state vector of the whole system, by using the fully quantum mechanical approach.
 For this purpose, all interactions between subsystems should be recognized and then, with the help of Schr\"{o}dinger equation or other appropriate methods, the state of the entire system may be found.
 Let us consider a model in which the single-mode quantized field which oscillates with frequency $\Omega$ in an optical cavity interacts with a three-level atom that is in the $\Lambda$-type atomic configuration. In this atomic configuration in which, the levels of the atom indicates by $|j\rangle$ with energies $\omega_{j}$, $ j=1,2,3$ (see Fig. 1), the transitions $|1\rangle\rightarrow|2\rangle$ and $|1\rangle\rightarrow|3\rangle$ are allowed whereas the transition $|2\rangle\rightarrow|3\rangle$ is forbidden in the electric-dipole approximation \cite{zubairy}. Also, we assume that the atom moves in the cavity and the atom-field interaction depends on the intensity of light. It is valuable to mention that intensity-dependent coupling can be easily realized by, for instance, algebraic generalization of the bosonic operators using nonlinear coherent states approach \cite{manko}. This formalism has shown its ability in the description of the center of mass motion of a trapped and bichromatically laser-driven ion \cite{vogel}. Also, regarding the realization of atomic motion, there exist some experiments that are comparable to the interaction of an atom with an electromagnetic pulse \cite{experiment}, in which the interaction of an atom with cavity eigenmodes of different shape functions is studied. Anyway, keeping in mind the above discussions and entering them appropriately in the standard JCM, the Hamiltonian describing the dynamics of our above system in the RWA can be written as ($\hbar=c=1$):
 \begin{eqnarray}\label{hamiltoni}
 \hat{H} =\hat{H}_{0}+\hat{H}_{1}, \nonumber
 \end{eqnarray}
 where
 \begin{eqnarray}\label{hamiltonih0}
 \hat{H}_{0}&=&\sum_{j=1}^{3} \omega_{j}\hat{\sigma}_{jj}+ \Omega \hat{a}^{\dag} \hat{a},
 \end{eqnarray}
 and
 \begin{eqnarray}\label{hamiltonih1}
 \hat{H}_{1}=\lambda_{1}f_{1}(z)(\hat{R}\; \hat{\sigma}_{13}+\hat{\sigma}_{31}\hat{R}^{\dag})
 + \lambda_{2}f_{2}(z)(\hat{R}\;\hat{\sigma}_{12}+\hat{\sigma}_{21}\hat{R}^{\dag}),
 \end{eqnarray}
 where $\hat{\sigma}_{ij}$ is the atomic transition operator defined by $\hat{\sigma}_{ij}=|i\rangle \langle j|,(i,j=1,2,3),\hat{a}$ and $\hat{a}^{\dag}$ are respectively bosonic annihilation and creation operators of the field and the constants $\lambda_{i}$, $i=1,2$, determine the atom-field coupling constants. Also, the operators $\hat{R} = \hat{a} g(\hat{n})$ and $\hat{R}^{\dag} = g(\hat{n}) \hat{a}^{\dag}$, with $\hat{n} = \hat{a}^{\dag} \hat{a}$ as the number operator of harmonic oscillator, denote the nonlinear ($f$-deformed) annihilation and creation operators, respectively. Using the well-known Weyl-Heisenberg Lie algebra corresponding to the operators $\hat{a}$, $\hat{a}^{\dag}$, $\hat{n}$ and the unity operator $\hat{I}$, and the fact that the operator $\hat{n}$ commutes with arbitrary function of itself, $g(\hat{n})$, the following communication relations clearly hold:
 \begin{eqnarray}\label{vrrd2}
 \left[\hat{R}, \hat{R}^{\dag}\right] &=& (\hat{n}+1)g^2(\hat{n}+1)-\hat{n} g^2(\hat{n}), \nonumber \\
 \left[\hat{R},\hat{n}\right] &=& \hat{R}, \;\;\; \left[\hat{R}^{\dag},\hat{n}\right]=-\hat{R}^{\dag},
 \end{eqnarray}
 where $g(\hat{n})$ is considered to be a Hermitian operator-valued function responsible for the intensity-dependent atom-field coupling.
 The influence of atomic motion in the model has been entered by the shape function $f_{i}(z)$.
 It is worth to note that a deep view in relation (\ref{hamiltonih1}) shows that this Hamiltonian may be reconstructed by changing $\lambda_{i}$ to $\lambda_{i}f_{i}(z)g(\hat{n}), i=1,2$, when it is compared with the standard JCM, i.e. the atom-field coupling depends on the atomic motion (by the shape function) and intensity of light (sometimes it is called $``$nonlinear JCM$"$ \cite{buzek}).\\
 We restrict our studies in the $z$-axis direction so that only the $z$-dependence of the field-mode function would be necessary to take into account. This is consistent with respect to the cavity QED experiments.
 The atomic motion would be incorporated as
 \begin{equation}\label{flamb}
  f_{i}(z)\rightarrow f_{i}(vt),\;\;\;\;i=1,2,
 \end{equation}
 where $v$ denotes the atomic velocity. To make the latter discussion more convenient, one may define a $\mathrm{TEM_{mnp_{i}}}$ mode as \cite{joshi1,joshi2,schlicher,lamb}
 \begin{equation}\label{fz}
 f_{i}(z)=\sin(p_{i}\pi vt/L ),
 \end{equation}
 where $p_{i}$ represents the number of half-wavelengths of the field mode inside a cavity with a length $L$.
 In order to obtain the state vector of the system, it should be suitable to rewrite Hamiltonian (\ref{hamiltoni}) in the interaction picture which results in
 \begin{eqnarray}\label{h-int}
 \hspace{-2cm} V=\lambda_{1}f_{1}(z)\left( \hat{R}\; \sigma_{13}e^{i\Delta_{3}t}+\sigma_{31}\hat{R}^{\dag}e^{-i\Delta_{3}t} \right)
 +\lambda_{2}f_{2}(z)\left( \hat{R}\; \sigma_{12}e^{i\Delta_{2}t}+\sigma_{21}\hat{R}^{\dag}e^{-i\Delta_{2}t} \right),
 \end{eqnarray}
 where $\Delta_{2}$ and $\Delta_{3}$ are the detuning parameters have been defined as
 \begin{eqnarray}\label{delta}
 \Delta_{2}&=&(\omega_{1}-\omega_{2})-\Omega, \nonumber \\
 \Delta_{3}&=&(\omega_{1}-\omega_{3})-\Omega.
 \end{eqnarray}
 Now, for simplicity and without loss of generality, we consider the resonance case in which $\Delta_{2}=\Delta_{3}=0$.
 Also, we assume that $\lambda_{1}=\lambda_{2} \equiv \lambda$  and
  $f_{1}(z) =f_{2}(z) \equiv f(z)$.\\
 Let us consider the initial state of the whole system to be in the following form:
 \begin{eqnarray}\label{sayi}
 |\psi(0)\rangle_{\mathrm{A-F}}=|1\rangle \otimes \sum_{n=0}^{+ \infty} q_{n} |n \rangle = \sum_{n=0}^{+\infty}q_{n}|1,n\rangle, \;\;\;\;\;
 \end{eqnarray}
  where $q_{n}$ is the probability amplitude of the initial radiation field of the cavity. Using the standard techniques, it may be found that, by the action of the time evolution operator (with the Hamiltonian in (\ref{h-int})), on the initial state vector of the system in (\ref{sayi}), we arrive at the explicit form of the wave function as follows
 \begin{eqnarray}\label{say}
 \hspace{-2cm} |\psi(t)\rangle=\sum_{n=0}^{+\infty}q_{n}\Big[ A(n,t) |1,n\rangle+B(n+1,t)|2,n+1\rangle
 + C(n+1,t)|3,n+1\rangle \Big]
 \end{eqnarray}
 where $A,B$ and $C$ are the atomic probability amplitudes which are given by
 \begin{eqnarray}\label{saycoef}
 A(n,t)&=&\cos \left[ \sqrt{2} \lambda \Theta(t) \sqrt{n+1} g(n+1) \right], \nonumber \\
 B(n+1,t)&=& \frac{1}{i \sqrt{2}} \sin \left[ \sqrt{2} \lambda \Theta(t) \sqrt{n+1} g(n+1) \right], \nonumber \\
 C(n+1,t)&=& \frac{1}{i \sqrt{2}} \sin \left[ \sqrt{2} \lambda \Theta(t) \sqrt{n+1} g(n+1) \right],
 \end{eqnarray}
 with the following definition for $\Theta(t)$
 \begin{equation}\label{theta1}
 \Theta(t)=\int_{0}^{t}f(vt')dt'=\frac{L}{p \pi v}[1-\cos(p \pi vt/L)].
 \end{equation}
 It is now necessary to emphasize the fact that obtaining the state vector of the entire system which has been acquired in (\ref{say}) is basically dependent on the initial state of the field. However, one can set $q_{n}$ by arbitrary amplitude of the initial state of the field such as number, phase, coherent or squeezed state. However, since the coherent state is more accessible than other typical field states (recall that the laser field far above the threshold condition is known as a coherent state \cite{zubairy}), we shall consider the field to be initially in a coherent state
 \begin{eqnarray}\label{amplitude}
 |\alpha \rangle = \sum_{n=0}^{+\infty} q_{n} |n \rangle, \;\;\;\;q_{n} = \exp \left( -\frac{ |\alpha|^{2} }{2}\right) \frac{\alpha^{n}}{\sqrt{n!}},
 \end{eqnarray}
 in which $|\alpha|^{2}$ is exactly the mean photon number (intensity of light). \\
 Now, in calculating the equation (\ref{theta1}), we have used the shape function $f(vt)$ from (\ref{fz}). For a particular choice of the atomic motion, we consider the velocity of the atom as in Refs. \cite{liao1,liao2} by $v=\lambda L/\pi$ and hence $f(z)=\sin(p \lambda t)$. Consequently, $\Theta(t)$ becomes
 \begin{equation}\label{theta2}
 \Theta(t)=\frac{1}{p \lambda}[1-\cos(p \lambda t)].
 \end{equation}
 Inserting (\ref{theta2}) in the time-dependent coefficients (\ref{saycoef}), the explicit form of the state vector of the entire system would be deduced.
 Henceforth, we are able to study the nonclassical features of such a system in the continuation of the paper.

 %===================================================================================================================================
 \section{Quantum entanglement between subsystems}
 %====================================================================================================================================

 Quantum entanglement is one of the most striking characteristics of quantum mechanics which plays a key role in many of the interesting applications of quantum computation and quantum information. Also, it is one of the main parts for the execution
 of quantum information processing devices \cite{qipd}. Recently, the degree of entanglement between atom and field which arises from the quantum interaction between them in an optical cavity has been reported in \cite{us}. In order to understand the degree of entanglement, the entropy is a useful concept. Entropy has a central role in classical information theory ({\it Shannon's entropy}) and quantum information theory ({\it von Neumann's entropy}) which measures how much uncertainty exists in the state of a physical system \cite{chuang}. The entropy of the field is a criterion which displays the strength of entanglement in which; higher (lower) entropy means the greater (smaller) degree of entanglement. For our purpose, we use the linear or von Neumann reduced entropy \cite{pk1}. Before obtaining the field entropy, it is valuable to pay attention to the important theorem of Araki and Leib \cite{araki}. This theorem expresses that for the two-components of considered quantum system, the entropies are limited by the following triangle inequality
 \begin{eqnarray}\label{valen}
 |S_{A}(t)-S_{F}(t)|\leq S(t) \leq S_{A}(t)+S_{F}(t),
 \end{eqnarray}
 where here the subscripts $` `$A$"$ and $` `$F$"$ refer to the atom and the field, respectively and the total entropy of the atom-field
 system is denoted by S. As a result of this theorem, if at the initial time the field and the atom are in pure states, the total entropy of the system is zero and remains constant. This means that, if the system is initially prepared in a pure state, at any time $t>0$, the entropy of the field is equal to the atomic entropy \cite{phoenix}. So, instead of the evaluation of the field entropy, we can focus on the entropy of the atom to arrive at the degree of entanglement.
 According to the von Neumann entropy as a measure of entanglement, the entropy of the atom and the field are defined through the corresponding reduced density operator by
 \begin{eqnarray}\label{ventd}
 S_{A(F)}(t)=-\mathrm{Tr}_{A(F)} \left(\hat{\rho}_{A(F)}(t) \ln \hat{\rho}_{A(F)}(t) \right).
 \end{eqnarray}
 The reduced density matrix of the atom required for evaluating (\ref{ventd}) is given by
 \begin{eqnarray}\label{vrdma}
 \hat{\rho}_{A}(t)&=&\mathrm{Tr}_{F}\left(  |\psi(t) \rangle \langle \psi(t) |   \right)  \nonumber \\
 &\dot{=}&\left(
  \begin{array}{ccc}
    \rho_{11} & \rho_{12} & \rho_{13} \\
    \rho_{21} & \rho_{22} & \rho_{23} \\
    \rho_{31} & \rho_{32} & \rho_{33} \\
  \end{array}
 \right).
 \end{eqnarray}
 The matrix elements in (\ref{vrdma}) read, for instance, by
 \begin{eqnarray}\label{rho}
 \rho_{11}&=&\sum_{n=0}^{+\infty}q_{n}q^{*}_{n}A(n,t)A^{*}(n,t), \nonumber \\
 \rho_{12}&=&\sum_{n=0}^{+\infty}q_{n+1}q^{*}_{n}A(n+1,t)B^{*}(n+1,t), \nonumber \\
 \rho_{13}&=&\sum_{n=0}^{+\infty}q_{n+1}q^{*}_{n}A(n+1,t)C^{*}(n+1,t),... \nonumber
 \end{eqnarray}
 where in all of the above relations, $q_{n}$ is the probability amplitude of the initial radiation field defined by (\ref{sayi}), and $A,B$ and $C$ are the
 atomic probability amplitudes derived in (\ref{saycoef}). Hence, the entropy of the field or atom can be obtained by the following
 relation \cite{us,pk3}
 \begin{eqnarray}\label{sff}
 S_{F}(t)=S_{A}(t)=-\sum_{j=1}^{3}\xi_{j} \ln \xi_{j}
 \end{eqnarray}
 where $\xi_{j}$, the eigenvalues of the reduced atomic density matrix in (\ref{vrdma}) are given by Kardan's instruction as \cite{kardan}
 \begin{eqnarray}\label{ventkardan}
 \xi_{j}&=&-\frac{1}{3}\alpha_{1}+\frac{2}{3}\sqrt{\alpha_{1}^{2}-3\alpha_{2}}\cos\left[\beta+\frac{2}{3}(j-1)\pi \right], \nonumber \\
 \beta &=& \frac{1}{3}\cos^{-1}\left[ \frac{9\alpha_{1}\alpha_{2}-2\alpha_{1}^{3}-27\alpha_{3}}{2(\alpha_{1}^{2}-3\alpha_{2})^{3/2}}\right],
 \end{eqnarray}
 with
 \begin{eqnarray}\label{vzal}
 \hspace{-2cm} \alpha_{1}&\dot{=}&-\rho_{11}-\rho_{22}-\rho_{33}, \nonumber \\
 \hspace{-2cm} \alpha_{2}&\dot{=}&\rho_{11}\rho_{22}+\rho_{22}\rho_{33}+\rho_{33}\rho_{11} -\rho_{12}\rho_{21}-\rho_{23}\rho_{32}-\rho_{31}\rho_{13}, \nonumber \\
 \hspace{-2cm} \alpha_{3}&\dot{=}& -\rho_{11}\rho_{22}\rho_{33}-\rho_{12}\rho_{23}\rho_{31}-\rho_{13}\rho_{32}\rho_{21} +\rho_{11}\rho_{23}\rho_{32}+\rho_{22}\rho_{31}\rho_{13}+\rho_{33}\rho_{12}\rho_{21}.
 \end{eqnarray}
 Equation (\ref{sff}) determines the variation of the entropy of the atom or the field with time. In addition, by this equation the degree of entanglement between the atom and field is also determined, that is, the subsystems are disentangled (the system of atom-field is separable) if equation (\ref{sff}) tends to zero.\\
 Figure 2 shows the time evolution of the field entropy against the scaled time $\tau=\lambda t$ for initial mean number of photons fixed at $|\alpha|^{2} = 10$. These plots attempt to indicate the influences of intensity-dependent coupling (left plots) by considering some particular operator-valued functions and atomic motion together with field-mode structure (right plots) by considering different values of $p$ in the shape function $f(z)$. We have chosen the nonlinearity functions as $g(n)=1$ (no intensity dependence, figure 2(a)), $g(n)=L_{n}^{1}(\eta ^{2})\left[(n+1)L_{n}^{0}(\eta ^{2})\right]^{-1}$ (figure 2(b)) associated with the center of mass motion of trapped ion \cite{Matos1996}, $g(n)=1/\sqrt{n}$ (figure 2(c)) which has been introduced by Man'ko {\it et al} \cite{manko} (where the corresponding coherent states have been named by Sudarshan as harmonious states \cite{harmonios}) and $g(n)=\sqrt{n+\nu}$ (figure 2(d)) corresponding to the well-known P\"{o}schl-Teller potential \cite{pt1,pt2}.
 It may be understood that for fixed $p$, the intensity-dependent coupling has no outstanding effect on the amount of field entropy, when it is compared with the situation in which intensity-dependent coupling is absent.
 We end this section with mentioning that by an increase in $p$, the intervals of time in which, the entropy or consequently the entanglement between the atom and field remains nearly in its maximum value, will be shorter. In other words, the variations between maxima and minima values of entropy in large $p$ is faster when we compare them with small value of field-mode structure parameter $p$. Also, it is seen that an increase in $p$ may be caused to reduce the degree of entanglement between subsystems. In addition, it is observed that in general, the influence of the atomic motion is more visible than the effect of intensity-dependent coupling.

 %===================================================================================================================================
 \section{Position-momentum entropic uncertainty relation and entropy squeezing}
 %===================================================================================================================================

 Heisenberg {\it uncertainty principle} expresses that in quantum mechanics, two noncommuting observables cannot be simultaneously measured with arbitrary precision. This observation is an essential limitation that is related neither to imperfections of the existing real-life measuring devices nor to the experimental errors of observation \cite{heisenberg}. For example, for $x$ and $p$ as two observables, Heisenberg uncertainty principle leads to the well-known inequality as $\Delta x \Delta p \geq 1/2$, where $\Delta x$ and $\Delta p$ correspond to the variance of the Hermitian operators $\hat{x}$ and $\hat{p}$ (quadrature operators of the radiation field), respectively. \\
 It should be noticed that the variance, which is used to define some quantum-mechanical effects such as quadrature squeezing of quantum fluctuations and sub-Poissonian statistics, is not the only measure of quantum uncertainty, and sometimes the``entropy'' may be preferred  instead of the ``variance''.
 Or{\l}owski \cite{Orlowski} has shown that besides the fact that, the entropic uncertainty relation is stronger than the standard uncertainty relation, the entropy (of the single observable) can be utilized as a measure of squeezing of quantum fluctuations as the variance. \\
 Paying attention to the above explanations and following Shannon's ideas \cite{shannon}, one may define the entropies of position and momentum $E_{x} = - \int P(x) \ln P(x) dx$ and $E_{p} = - \int P(p) \ln P(p) dp$, where $P(x)$ and $P(p)$ are defined as $P(x) = \langle x|\hat{\rho}_{f}|x\rangle $ and $P(p) = \langle p|\hat{\rho}_{f}|p\rangle $, respectively. The sum of the above-mentioned position and momentum entropies leads to the interesting inequality which is given by \cite{Orlowski}
\begin{eqnarray}\label{ineqxp}
 E_{x}+ E_{p} \geq 1 + \ln \pi.
 \end{eqnarray}
This inequality is often called the position-momentum entropic uncertainty relation that has been proven by Beckner \cite{proof-EUR} for the first time. It is valuable to notice that the entropies of position and momentum corresponding to the standard coherent state (and to the vacuum state) are equal to each other, that is, $E_{x}=E_{p}=\frac{1}{2}(1+\ln \pi)$ \cite{Orlowski}.
Considering the latter inequality, one can supply an alternative mathematical formulation of the uncertainty principle by the inequality \cite{ERspringer,honarasa1}
 \begin{equation}\label{deltaxp}
 \delta x \delta p \geq \pi e,
 \end{equation}
 where $\delta x $ and $\delta p $ are defined as the exponential of Shannon entropies associated with the probability distributions for $x$ and $p$ which are given by
 \begin{eqnarray}\label{shen}
 \delta x &=&\exp (E_{x})=\exp\left(-\int_{-\infty}^{+\infty}\langle x|\hat{\rho}_{f}|x\rangle \ln \langle x|\hat{\rho}_{f}|x\rangle dx \right), \nonumber \\
 \delta p &=&\exp (E_{p})=\exp\left(-\int_{-\infty}^{+\infty}\langle p|\hat{\rho}_{f}|p\rangle \ln \langle p|\hat{\rho}_{f}|p\rangle dp \right),
 \end{eqnarray}
 where the density matrix element may be determined as follows
 \begin{eqnarray}\label{xrx}
 \langle x|\hat{\rho}_{f}|x\rangle &=& \left| \sum_{n=0}^{+\infty} q_{n} A(n,t) \langle x|n\rangle\right|^{2}
 +\left| \sum_{n=0}^{+\infty} q_{n} B(n+1,t) \langle x|n+1\rangle\right|^{2} \nonumber \\
 &+& \left| \sum_{n=0}^{+\infty} q_{n} C(n+1,t) \langle x|n+1\rangle\right|^{2},
 \end{eqnarray}
 in which
 \begin{equation}\label{xn}
 \langle x|n\rangle=\left[ \frac{\exp\left(-x^{2}\right)}{\sqrt{\pi}2^{n}n!} \right]^{1/2} H_{n}(x),
 \end{equation}
 with $H_{n}(x)$ as the Hermite polynomials. It is worthwhile to mention that entropic uncertainty relations such as equation (\ref{deltaxp}) do physically imply the fact that having complete information about the values of a pair of observables with no common eigenstates is {\it simultaneously} impossible \cite{Ruiz}. \\
 Now, in order to analyze the entropy squeezing properties of the atom in motion, we introduce two normalized quantities
 \begin{eqnarray}\label{quan}
 \mathbf{E}_{x}(t)&=&(\pi e)^{-1/2}\exp(E_{x}(t))-1, \nonumber \\
 \mathbf{E}_{p}(t)&=&(\pi e)^{-1/2}\exp(E_{p}(t))-1.
 \end{eqnarray}
 When $-1<\mathbf{E}_{x}(t)<0$ ($-1<\mathbf{E}_{p}(t)<0$), the position (momentum) or $x (p)$ component of the field entropy is said to be squeezed.\\
 Presented results in figure 3 show the time evolution of entropy squeezing versus the scaled time $\tau$ for the initial mean number of photons fixed at $|\alpha|^{2} = 10$. Figure 3(a) is plotted for the situation in which no intensity-dependent coupling exists. This plot shows that the entropy squeezing changes with scaled time between positive and negative values and in some intervals of time, squeezing occurs. Adding the effect of the intensity dependence by some special nonlinearity functions, which have been displayed in the related plots, it is seen from figures 3(b), 3(c) and 3(d) that squeezing occurs in more long intervals of time in comparison with figure 3(a) and according to figure 3(c), squeezing appears for all the time. We conclude that intensity-dependent coupling has a significant role in increasing the negativity of entropy squeezing in position component. Also, the right plots of figure 3, in which the effect of atomic motion and field-mode structure is studied, show that depending on the nonlinearity function $g(n)$, the negativity of entropy squeezing may be decreased, when the value of $p$ grows.

 %===================================================================================================================================
 \section{Photon statistics: Mandel parameter}
 %===================================================================================================================================

 To characterize the statistical properties of the field, the parameter that has been frequently used, is the Mandel parameter, which measures the deviation from Poissonian distribution. This parameter has been defined as follows \cite{mandel}
 \begin{eqnarray}\label{mandel1}
 Q = \frac{\langle (\hat{a}^{\dag}\hat{a})^2\rangle - \langle \hat{a}^{\dag}\hat{a} \rangle^{2}}{\langle \hat{a}^{\dag}\hat{a} \rangle}-1.
 \end{eqnarray}
 This quantity vanishes for $` `$standard coherent light$"$ (Poissonian), is positive for $` `$classical$"$ or $` `$chaotic light$"$ (supper-Poissonian), and negative for $` `$nonclassical$"$ light$\;$(sub-Poissonian).
 For current formalism we have
 \begin{eqnarray}\label{n}
 \hspace{-2cm} \langle \hat{a}^{\dag}\hat{a} \rangle&=&\sum_{n=0}^{+\infty}P_{n}\Big[n|A(n,t)|^{2}+(n+1) \big(|B(n+1,t)|^{2}
 +|C(n+1,t)|^{2} \big)\Big],
 \end{eqnarray}
 and in a similar manner
 \begin{eqnarray}\label{n2}
 \hspace{-2cm} \langle (\hat{a}^{\dag}\hat{a})^{2} \rangle&=&\sum_{n=0}^{+\infty}P_{n}\Big[n^{2}|A(n,t)|^{2}+(n+1)^{2} \big(|B(n+1,t)|^{2}
 +|C(n+1,t)|^{2} \big)\Big],
 \end{eqnarray}
 in which, probability amplitudes $ A $, $ B $ and $ C $ have been found in (\ref{saycoef}).\\
 Our presented results in figure 4 indicate the effects of intensity-dependent coupling (left plots) and atomic motion together with the field-mode structure (right plots) on the time evolution of Mandel parameter versus the scaled time $\tau$ for the initial mean number of photons fixed at $|\alpha|^{2} = 10$. Figure 4(a) corresponds to the situation in which intensity-dependent coupling is absent and the other plots (figures 4(b), 4(c) and 4(d)) study the effect of the presence of intensity-dependent coupling. According to these plots, intensity-dependent coupling has no striking enhancing role on the negativity of Mandel parameter and can bring about a decrease in the amount of the negativity of this nonclassicality indicator depending on the chosen nonlinearity function which is selected. Also, the right plots of figure 4, which refer to the effect of atomic motion and field-mode structure, indicate that for some nonlinearity functions (figures 4(a) and 4(c)), the increase of parameter $p$ may reduce the negativity of Mandel parameter.

 %===================================================================================================================================
 \section{Quadrature (normal) squeezing}
 %===================================================================================================================================

 In quantum optics, squeezing property is described by the reduction of quantum fluctuation in one of the field quadratures below its value of vacuum or canonical coherent states. Considering the following Hermitian operators $\hat{x}=(\hat{a}+\hat{a}^{\dag})/2$ and $\hat{p}=(\hat{a}-\hat{a}^{\dag})/2i$, the $\hat{x}$ and $\hat{p}$ quadratures obey the commutation relation $[\hat{x},\hat{p}]=i/2$.
 Consequently, the uncertainty relation for such operators reads as $\left(\Delta \hat{x}\right)^{2}\left(\Delta \hat{p}\right)^{2}\geq 1/16$, where $\langle\Delta \hat{z} \rangle ^{2}=\langle \hat{z}^{2} \rangle-\langle \hat{z} \rangle^{2}$ and $\hat{z}=\hat{x}\;\mathrm{or} \; \hat{p}$ and $\Delta \hat{x}$ and $\Delta \hat{p}$ are the uncertainties in the quadrature operators $\hat{x}$ and $\hat{p}$, respectively.
 A state is squeezed in $\hat{x} \;(\hat{p})$ if $\left(\Delta \hat{x}\right)^{2}<0.25\; (\left(\Delta \hat{p}\right)^{2}<0.25)$, or equivalently by defining the variation (squeezing) parameters
 \begin{eqnarray}\label{vsqdx1}
 \hat{V}_{x}=4 \left(\Delta \hat{x}\right)^{2} -1, \hspace{1cm} \hat{V}_{p}=4 \left(\Delta \hat{p}\right)^{2} -1,
 \end{eqnarray}
 squeezing occurs in $\hat{x} \;(\hat{p})$ component respectively if $-1<\hat{V_{x}}<0 \;(-1<\hat{V_{p}}<0) $.
 These parameters can be rewritten as
 \begin{eqnarray}\label{vsqx1}
 \hat{V}_{x}&=&2\langle \hat{a}^{\dag} \hat{a} \rangle + \langle \hat{a}^{2} \rangle + \langle \hat{a}^{\dag 2} \rangle-\left( \langle \hat{a} \rangle + \langle \hat{a}^{\dag} \rangle \right)^{2}, \nonumber\\
 \hat{V}_{p}&=&2\langle \hat{a}^{\dag} \hat{a} \rangle - \langle \hat{a}^{2} \rangle - \langle \hat{a}^{\dag 2} \rangle+\left( \langle \hat{a} \rangle - \langle \hat{a}^{\dag} \rangle \right)^{2},
 \end{eqnarray}
 where the necessary expectation value of photon number has been obtained in (\ref{n}) and the following general relation
 \begin{eqnarray}\label{ar}
 \hspace{-2cm} \langle \hat{a}^{r}\rangle&=&\sum_{n=0}^{+\infty}q_{n}^{*}q_{n+r}\Bigg[\sqrt{\frac{(n+r)!}{n!}} A^{*}(n,t)A(n+r,t) \nonumber \\
 \hspace{-2cm} &+& \sqrt{\frac{(n+r+1)!}{(n+1)!}}\bigg( B^{*}(n+1,t)B(n+1+r,t)
 +  C^{*}(n+1,t)C(n+1+r,t)\bigg)\Bigg],
 \end{eqnarray}
 where $ A $, $ B $ and $ C $ have been determined in (\ref{saycoef}) and $\langle \hat{a}^{r} \rangle ^{*}=\langle \hat{a}^{\dag\;^{r}} \rangle$.\\
 Figure 5 studies the influences of intensity-dependent coupling and atomic motion on the first-order (normal) squeezing in $\hat{x}$ quadrature in terms of scaled time for different chosen parameters similar to figure 2.
 Figure 5(a) has been depicted in the absence of intensity-dependent coupling and indicates oscillatory behaviour between positive and negative values and so the state of the system is squeezed in some intervals of time. The plots 5(b), 5(c) and 5(d) are related to the situation in which, the influence of intensity-dependent coupling has been entered. These plots show that the amount of quadrature squeezing can be increased/decreased by entering the effect of intensity-dependent coupling and depending on choosing the nonlinearity function, this property may be improved. Although, according to the right plots of figure 5, with increasing the field-mode structure parameter $p$, the amount of negativity of normal squeezing may be descended.

 %============================================================================================================

 \section{Summary and concluding remarks}\label{examples}

 %===================================================================================================================================

 In this paper, we have studied the nonlinear interaction between a moving $\Lambda$-type three-level and a single-mode cavity field
 using the generalized JCM. In a sense, our formalism is based on transforming the coupling constant $\lambda$ to velocity- and intensity-dependent coupling $\lambda f(vt)g(n)$. Next, after obtaining the explicit form of the state vector of the whole atom-field system, quantum entanglement between the subsystems has been investigated by using the von Neumann reduced density matrix approach. Also, by considering the Shannon's idea, the position-momentum entropic uncertainty relation is considered, from which the entropy squeezing of the state vector of the entire system has been numerically examined.
 In addition, two nonclassicality features namely quadrature squeezing of the field and Mandel parameter, as the well-known and important nonclassical criteria, have been studied.
 Particularly, the effect of $` `$intensity-dependent coupling$"$ by considering a few nonlinearity functions associated with different physical systems such as $g(n)=1$ (no intensity dependence), $g(n)=L_{n}^{1}(\eta ^{2})\left[(n+1)L_{n}^{0}(\eta ^{2})\right]^{-1}$ (associated with the center of mass motion of trapped ion), $g(n)=1/\sqrt{n}$ (corresponding to the harmonious state) and $g(n)=\sqrt{n+\nu}$ (which refers to the P\"{o}schl-Teller potential) has been illustrated. Also, the influence of $` `$atomic motion$"$ has been illustrated by considering different values of field-mode structure $p$.\\
 It has been observed that the intensity-dependent coupling has no remarkable effect on the degree of entanglement between the atom and the field, when it is compared with the situation in which intensity-dependent coupling does not exist. Altogether, we observed that intensity-dependent coupling has a significant role in increasing the negativity of entropy squeezing in position component. Also, intensity-dependent coupling has no striking effect on the negativity of Mandel parameter and indeed, the strength of sub-Poissonian statistics may be noticeably reduced by choosing the special nonlinearity function (figure 4c). In addition, it is seen that entering the effect of intensity-dependent coupling can ascend/descend the amount of quadrature squeezing depending on the chosen nonlinearity function $g(n)$.\\
 Now, we pay attention to the effect of atomic motion on the considered nonclassicality criteria. From the obtained results, it is observed that in all related figures with different nonlinearity functions $g(n)$, increasing the value of field-mode structure parameter $p$ follows generally by decreasing the ``range'' of the nonclassicality signs. Also, the reduction of the ``amount'' of these nonclassicality features may be arisen from an increase in the value of parameter $p$. \\
 Consequently, according to our numerical results, we can deduce that the duration of maximum amount of each of the nonclassicality indicators can be controlled by choosing the appropriate nonlinearity function and the field-mode structure parameter $p$, when initial states of the subsystems are fixed.\\
 Moreover, the atomic motion together with the nonlinearity function leads to the oscillatory and periodic behaviour of the degree of entanglement between the atom and field and nonclassical properties in contrast to the situation in which the atomic motion is absent and the atom-field coupling is constant ($g(n)=1$) that a chaotic behaviour for the time evolution of the field entropy and nonclassical features is observed \cite{us}.\\
 Finally, we would like to emphasize the generality of our model in the sense that it may be used for any physical system, either any nonlinear oscillator with arbitrary nonlinearity function $g(n)$, or any solvable quantum system with known $e_{n}$, using the relation $e_{n}=ng^{2}(n)$ \cite{en}. This study can be performed by considering different configurations of three--level atom ($\Xi$-- and $V$--type configurations) and/or different preparations for the initial state of the atom and/or the field, too.

%=============================================================================================================
%=============================================================================================================
\begin{flushleft}
 {\bf Acknowledgements}\\
 \end{flushleft}
 The author would like to express their utmost thanks to the referees for their intuitive comments which improved the paper.

 %======================
%==============================================
 %\newpage

 %\section*{References}

 \end{document}